\documentclass[epsf,prb,twocolumn,showpacs,nofootinbib]{revtex4-1}

\usepackage[pdftex]{graphicx}
\usepackage{dcolumn}
\usepackage{bm}
\usepackage{epsfig}
\usepackage{latexsym}
\usepackage{amsmath}
\usepackage{amsfonts}
\usepackage{amssymb}
\usepackage{color}
\usepackage{array}
\usepackage{framed}

\begin{document}

\title{Subdimensional particle structure of higher rank $U(1)$ spin liquids}
\author{Michael Pretko\\
\emph{Department of Physics, Massachusetts Institute of Technology,
Cambridge, MA 02139, USA}}
\date{March 5, 2017}

\begin{abstract} 
Spin liquids are conventionally described by gauge theories with a vector gauge field.  However, there exists a wider class of spin liquids with higher rank tensors as the gauge variable.  In this work, we focus on (3+1)-dimensional spin liquids described by $U(1)$ symmetric tensor gauge theories, which have recently been shown to be stable gapless spin liquids.  We investigate the particle structure of these tensor gauge theories and find that they have deep connections with the ``fracton" models recently discovered by Vijay, Haah, and Fu.  Tensor gauge theories have more conservation laws than the simple charge conservation law of rank 1 theories.  These conservation laws place severe restrictions on the motion of particles.  Particles in some models are fully immobile (fractons), while other models have particles restricted to motion along lower-dimensional subspaces.
\end{abstract}
\maketitle

\normalsize

\section{Introduction}

Spin liquids are strongly-interacting spin systems which exhibit long-range entanglement in the ground state and an exotic excitation spectrum which is not smoothly connected to the non-interacting limit.  See Reference \onlinecite{lucile} for an entanglement-based review of spin liquids.  Theoretically, spin liquids correspond to the deconfined phases of compact gauge theories, and we can mostly use the terms spin liquid and gauge theory interchangeably.\cite{foot1}  In the gapped case, an equivalent concept is topological order.\cite{xgtop}  We use the term spin liquid in a more general sense, referring to stable phases with long-range entanglement, either gapped or gapless, all described quite naturally as deconfined gauge theories.  To date, the study of spin liquids has mostly focused on rank 1 gauge theories, by which we mean the gauge variable $A_i$ (``vector potential") is a vector, $i.e.$ a rank 1 tensor.  This gauge variable can take discrete or continuous values, or even be matrix valued in the case of non-abelian gauge theories.

In the present work however, we will investigate a class of spin liquids going beyond this traditional paradigm.  While the vector potential theory is most familiar, there is in general no reason why the emergent gauge variable should be a vector, as opposed to a more complicated geometric object, such as a higher rank tensor.  (The meaning of ``tensor" will be discussed further in the Appendix.)  We therefore wish to explore the consequences of a tensor gauge structure, to see whether or not we get any exotic new physics.  We will briefly discuss later the prospects of realizing such phases in experiments, but ultimately the motivation here will not come from a specific physical system.  Rather, we are seeking to expand the boundaries of what sorts of emergent behavior can occur in highly entangled phases of matter.  We will find that the phases considered here represent a drastic departure from the properties of rank 1 spin liquids and represent an exciting new field of study.

As a first thought on going beyond the vector framework, one might consider working in the language of differential forms.  For example, while a vector potential $A_i$ naturally describes a phase of condensed strings ending on point-like excitations, the antisymmetric rank 2 tensor $A_{ij}$ (``two-form") naturally describes a phase of condensed surfaces terminating on string-like excitations.  However, antisymmetric tensor theories do not lead to new stable spin liquids in 3+1 dimensions.  In the discrete case, the two-form simply leads to a dual description of the conventional one-form, swapping the role of the electric particles and magnetic loops.  In the $U(1)$ case, the two-form (called a Kalb-Ramond field in the string theory literature) actually is distinct from the one-form, but unfortunately these models are unstable to gapped confined phases in 3+1 (or fewer) dimensions \cite{analytic,numerical,soojong,gerbe}, so they do not correspond to realistic spin liquids in our world.\cite{foot2}  We also expect non-abelian generalizations to share this instability to confinement, so pure antisymmetric tensor gauge theories seem to contain no new physics in 3+1 dimensions.

On the other hand, the case of symmetric tensor $U(1)$ gauge theories\cite{foot3} has recently been analyzed by Rasmussen, You, and Xu\cite{alex}, who showed that there are multiple theories of a given rank, and all such theories are stable gapless phases in 3+1 dimensions (but unstable in 2+1).  Furthermore, there is reason to be optimistic that these phases can be found in realistic spin systems, as we shall comment on further in the conclusion.  We shall therefore focus on (3+1)-dimensional symmetric tensor theories in this work, though a brief mention will be made of mixed-symmetry tensors, which may or may not be stable.  (Note that a general tensor gauge theory does not fall into the category of ``higher form" gauge theory, which refers specifically to antisymmetric tensor objects.  A generic tensor does not correspond to a differential form.)  The symmetric tensor gauge theories represent a new class of stable gapless spin liquids, where the gaplessness is protected against any microscopic perturbation to the Hamiltonian, without regards to symmetry.  However, important questions remain.  How do we know that these theories are really distinct from previously understood spin liquids, and are not simply dual formulations of some rank 1 theory, or perhaps several rank 1 theories stapled together?  Is there some property of these theories which fundamentally distinguishes them from the rank 1 case?  We shall here answer this question in the affirmative for the $U(1)$ case by examining the particle structure of higher rank $U(1)$ spin liquids.  It will be found that, though the gapless gauge modes of these systems are qualitatively similar to the rank 1 case, the particle structure is dramatically different.

The particles in these theories have severe restrictions placed on their motion, in a manifestation of a phenomenon encountered earlier in the condensed matter literature.  This sort of restricted mobility was first seen in work due to Chamon\cite{chamon} and in several other models\cite{bravyi,cast,yoshida}, including the famous Haah's code\cite{haah,haah2}.  These principles were later more systematically developed in the ``fracton" theories of Vijay, Haah, and Fu \cite{fracton1,fracton2}.  (We also refer the reader to other recent literature on the fracton phenomenon.\cite{williamson,sagarlayer,hanlayer,abhinav})  Depending on the gauge constraint, particles in tensor gauge theories can be restricted to motion along lower-dimensional subspaces, or they can be prevented from moving at all without the creation of other excitations.  Following the conventions of Reference \onlinecite{fracton1}, the former will be referred to as $d$-dimensional particles, where $d$ is the dimension of the subspace, and the latter will be called fractons.  We will use the phrase ``subdimensional particles" as a blanket term for both, since fractons can naturally be viewed as $0$-dimensional particles.

The restriction of charge motion in all of these models has a very natural interpretation in terms of higher-moment conservation laws.  For example, whereas the rank 1 spin liquids only had a charge neutrality constraint, we will see that particle configurations in a rank 2 spin liquid with scalar charge must have both charge neutrality and vanishing dipole moment.  Other higher rank theories can be understood in terms of similar conservation laws.  These extra conservation laws lead immediately to subdimensional behavior in these models.

While all of these models are stable against the usual Polyakov mechanism for confinement\cite{polyakov}, where monopoles proliferate and gap the gauge field, we find that a subset of these models exhibit a new, milder form of ``confinement," which we term electrostatic confinement, arising from the large energy stored in the electric field of a static point charge.  This mechanism involves a large energy cost for widely separated particles, but it does not gap the gauge field.  And in many cases, surprisingly, it still leaves the particles as well-defined excitations, so most of the usual properties of deconfined phases are still present.  In other cases, however, electrostatic confinement does destroy the charge sector of the theory, but subdimensional neutral bound states are still possible and the gapless sector remains untouched, so these theories are still far from trivial.  This unconventional confinement mechanism is closely connected with the restricted motion of fractons.

The models here represent gapless generalizations of the original discrete fracton models.  It seems clear that, by considering discrete symmetric tensor gauge theories, such as by breaking the $U(1)$ theories down to a discrete subgroup, we could generate a large class of gapped models with subdimensional particles.  It seems likely that the ``generalized gauge theory" of Reference \onlinecite{fracton2} can be naturally phrased in terms of a higher rank gauge theory (not necessarily symmetric).  It will be an interesting future project to see if all of the fracton models can be described in terms of higher rank gauge theories, and to investigate what the precise relation is between higher rank gauge theory and the techniques from algebraic geometry used in the original fracton constructions.

We also note that the higher rank gauge theories described here might have been called ``higher spin" gauge fields by our high energy colleagues.  (We have avoided this terminology, since we have abused the term spin enough already.)  Fields of spin higher than 2 receive much less attention in the high energy literature than their lower spin counterparts, since it is hard to construct consistent interacting theories.  For many years, the only consistently interacting higher spin framework has been Vasiliev theory\cite{vasiliev} and its variants, which involve an infinite tower of all possible higher spin gauge fields.  In light of the present work, it would seem that individual higher spin gauge fields actually can be consistently coupled to matter, as long as the matter particles have subdimensional behavior.  The difficulties which have plagued the development of higher spin gauge fields in the high energy literature seem to be a manifestation of the fact that such fields cannot be consistently coupled to fully mobile matter fields.

\section{Review of the Rank 1 $U(1)$ Spin Liquid}

The models that we will consider in this paper are natural extensions of the conventional rank 1 $U(1)$ spin liquid in three spatial dimensions.  It will therefore be useful to first review the basic structure of this more familiar theory.  In such a theory, we can take our basic degrees of freedom to be those of a vector field $A_i$.  We let the theory be compact, so that the components of $A_i$ are $U(1)$-valued.  In other words, we identify $A_i\sim A_i + 2\pi$, which effectively makes each component a quantum rotor variable.  We will also work with the canonical conjugate of the gauge field, which is the electric field $E_i$, as is familiar from classical electromagnetism.  $E_i$ effectively serves as the angular momentum of the rotor variable and as such is quantized to have integer eigenvalues.  In a concrete lattice model, $A_i$ and $E_i$ are most conveniently taken to live on links of the lattice.  The choice of lattice will not be important for the essential physics, though oftentimes the presence of a lattice makes issues like compactness easier to deal with.

We now seek a low-energy theory which has the properties of the familiar Maxwell theory of a $U(1)$ gauge field.  Motivated by this, we often characterize a $U(1)$ spin liquid by demanding that the low-energy theory be invariant under the gauge transformation $A_i(x)\rightarrow A_i(x) + \partial_i \alpha(x)$, for a function $\alpha(x)$ with arbitrary spatial dependence.  For our purposes, however, it is actually more convenient to work in terms of the canonically conjugate variable $E_i$.  For a state to be invariant under $A_i\rightarrow A_i + \partial_i \alpha$, it is easy to show that the state must obey the constraint $\partial_i E^i = 0$, which is simply the source-free Gauss's law.  It is also easy to show that the low-energy theory consistent with this gauge constraint/transformation takes the form:
\begin{equation}
H =\frac{1}{2}g\sum_{links} E^2 - \sum_{plaquettes}\cos B
\label{lowen}
\end{equation}
where $B_i = \epsilon_{ijk}\partial^jA^k$ is the magnetic field (conventionally defined on plaquettes of the lattice), and $g$ is a tuning parameter.  When $g$ is small, the fluctuations of the cosine around its minimum will be small, and we can write:
\begin{equation}
H\rightarrow\int d^3x\frac{1}{2}(gE^2 + B^2)
\end{equation}
which should look familiar from electromagnetism.  This Hamiltonian leads to a gapless photon mode with linear dispersion and two polarizations, as we expect.  At large $g$, however, the electric term dominates, effectively disordering the magnetic cosine term, and the system picks up a gap of order $g$.  All of the properties of Maxwell theory are destroyed, and the system will no longer be in a spin liquid phase.  This is the phenomenon of confinement.

At low energies, we have characterized the $U(1)$ spin liquid by a gapless photon mode whose field configurations obey the constraint $\partial_i E^i = 0$.  However, there is another important set of excitations to consider.  While the low-energy sector obeyed a constraint, there can be states higher up in energy which violate this constraint, $\partial_i E^i = \rho\neq 0$.  These violations of the low-energy constraint correspond precisely to the charges of the emergent electric field, which must be present on very general grounds.  If we demand that our spin liquid exist within a tensor product Hilbert space (as any real spin liquid does), then the existence of emergent charges is guaranteed \cite{me}.  Furthermore, this charge represents a conserved quantity, as no local operator can create a net charge in the system.  If we work on a closed manifold, then we are always guaranteed to have no net charge in the system:
\begin{equation}
\int \rho = \int \partial_iE^i = 0
\end{equation}
since we are integrating a total derivative.  In general, these charges will exist as gapped excitations of the system and will have a corresponding energy penalty in the Hamiltonian.  While Equation \ref{lowen} was valid for the gapless sector, we will more generally have:
\begin{equation}
H =\frac{1}{2}g\sum_{links} E^2 - \sum_{plaq}\cos B + U\sum_{vert}(\partial_iE^i)^2 + \cdot\cdot\cdot
\label{genham}
\end{equation}
where the $U$ term is defined on each vertex of the lattice.  The ``$\cdot\cdot\cdot$" represents the terms which do not commute with the gauge constraint, which are irrelevant to the low-energy physics.  These ``$\cdot\cdot\cdot$" terms are important, however, as they will dictate the dynamics of the charges.

There is also one last class of excitations to consider: magnetic monopoles.  The magnetic field $B_i = \epsilon_{ijk}\partial^j A^k$ seems to automatically obey $\partial_iB^i = 0$.  However, we must remember that $A_i$ is only defined modulo $2\pi$, which allows us to more generally have $\partial_iB^i = 2\pi n$, for integer $n$.  Such a defect corresponds to a magnetic charge of strength $n$.  The properties of these excitations closely mirror those of the analogous electric particles.  This is a manifestation of electromagnetic duality.  By making use of the constraint $\partial_i E^i = 0$, we can write $E^i = \epsilon^{ijk}\partial_j \tilde{A}_k$ within the low-energy sector, for a dual gauge potential $\tilde{A}$.  In terms of this variable, the electric and magnetic fields have simply switched places, giving us a useful dual formulation of the theory.

The most important fact about the magnetic monopoles in this three-dimensional system is that they are excitations with a well-defined energy.  We can therefore imagine a phase in which the monopoles are gapped and do not effect the low-energy physics.  This allows the $U(1)$ spin liquid to exist as a stable phase of matter.  This is in sharp contrast to the two-dimensional case, in which magnetic monopoles are ``instantons" (spacetime defects) corresponding to phase slip events, not excitations.  There is no meaningful sense in which instantons can be gapped, so there is no guarantee that they will not effect the low-energy physics.  Indeed, it was shown by Polyakov\cite{polyakov} that the two-dimensional $U(1)$ gauge theory is totally destroyed by these instantons.  There do exist schemes for stabilizing a two-dimensional $U(1)$ spin liquid against instantons, such as by coupling to a large number of gapless charges\cite{stability}, but this is a more complicated problem.  In this work, we will only consider theories which are instanton-free and are therefore stable.

\section{Rank 2 Theories}

We now proceed to the higher rank theories.  We will first go through the rank 2 case to illustrate the principle, which is readily generalized.  We take our degrees of freedom to be those of a compact $U(1)$-valued symmetric tensor $A_{ij}$, with canonically conjugate variable $E_{ij}$, representing a generalized electric field.  Each component $A_{ij}$ is essentially a quantum rotor, and the momentum variable $E_{ij}$ is quantized to have integer values.  In the simplest lattice models\cite{cenke2}, off-diagonal elements ($e.g.$ $E_{xy}$) naturally live on faces of a lattice, while diagonal elements ($e.g.$ $E_{xx}$) naturally live on vertices, but the precise choice of lattice system will not be important for the discussion here.  As discussed in Reference \onlinecite{alex}, there are three different sorts of gauge transformations of $A_{ij}$ which can be considered at rank 2, leading to three different analogues of Gauss's law which one can write down: $\partial_i E^{ij} = 0$, $\partial_i\partial_j E^{ij} = 0$, and $E^i_{\,\,i} = 0$, some of which can be applied on top of each other.  Furthermore, each valid combination of Gauss's laws will represent a stable phase, as we will discuss.

To construct Hamiltonians for these theories, the authors of Reference \onlinecite{alex} considered the natural generalization of the rank 1 compact U(1) Hamiltonian, Equation \ref{genham}.  The generalized $E$ term and gauge constraint term can be written down immediately, whereas the generalized $B$ term requires a bit more cleverness.  The structure of the $B$ term depends on the gauge constraint, and there can be different numbers of spatial derivatives in $B$ depending on the theory, leading to different dispersions for the gauge mode.  We will delay discussion of the magnetic tensor until a later section, since most of our analysis will not need to make any use of the specifics of these Hamiltonians, except for the $U$ term enforcing the generalized Gauss's law.  Almost all of the important physics follows directly from the Gauss's law.  The other terms in the Hamiltonian only serve to define the dynamics of the gapless gauge mode and the structure of the magnetic defects.  We will, of course, need to check later that these magnetic defects are not instantons, so that the theory is stable.  This will indeed be the case, so that these phases will all have a stable deconfined phase at small $g$ (and obviously a trivial confined phase at large $g$).  As first shown in Reference \onlinecite{alex}, many of the models considered in this paper will have an electric-magnetic duality, so the behavior of the magnetic particles will often be the same as that of the electric particles, which we focus on first.  All we will need for the present discussion is that the gauge field is not confining\cite{foot4} at small $g$, so that particles exist as well-defined excitations in this phase.

It is also worth noting that we expect such rank 2 symmetric tensor theories to have some relationship with the theory of gravity, which is also described by a symmetric tensor gauge field.  There is actually a deep connection between the models considered here and emergent gravity, but this relationship will not be apparent at the level of the analysis we will conduct here.  The emergent gravitational behavior of these phases is a topic of its own and is being treated in a separate work.\cite{mach}

\subsection{Scalar Charge Theory}

Let us first take the example of imposing only the constraint $\partial_i\partial_j E^{ij} = 0$, corresponding to the gauge transformation $A_{ij} \rightarrow A_{ij} + \partial_i\partial_j \phi$ for arbitrary scalar function $\phi$.  Of course, the source-free gauge constraint applies only to the low-energy subspace, achieved for example via a term in the Hamiltonian of the form $U(\partial_i\partial_j E^{ij})^2$ for large $U$.  States which violate the source-free Gauss's law must appear higher up in energy as particle states of the theory in order to have a tensor product Hilbert space, as is the situation in any condensed matter problem (see Reference \onlinecite{me} for further discussion of this issue).  For a general state, we can therefore write the generalized Gauss's law as $\partial_i\partial_j E^{ij} = \rho$, defining $\rho$ as the scalar charge density.

So what conservation laws do we have in this system?  Obviously we have charge neutrality, just as in the rank 1 case:
\begin{equation}
\int \rho = \int \partial_i\partial_j E^{ij} = 0
\end{equation}
where the integrals are over three-dimensional space, and we have integrated a total derivative term.  (We choose to work on a closed manifold for simplicity, so that the integral of the total derivative vanishes.  Everything works similarly on an open manifold.)  This conservation law leads to the usual constraint that the emergent charges cannot be created or destroyed unless it is accompanied by the creation/destruction of other charges in order to preserve neutrality.  Naively, one such allowed neutrality-preserving operation is a local hop: a particle is destroyed on one site and created on a neighboring site, in accordance with our usual intuition of particle mobility.

\begin{figure}[t!]
 \includegraphics[scale=0.35]{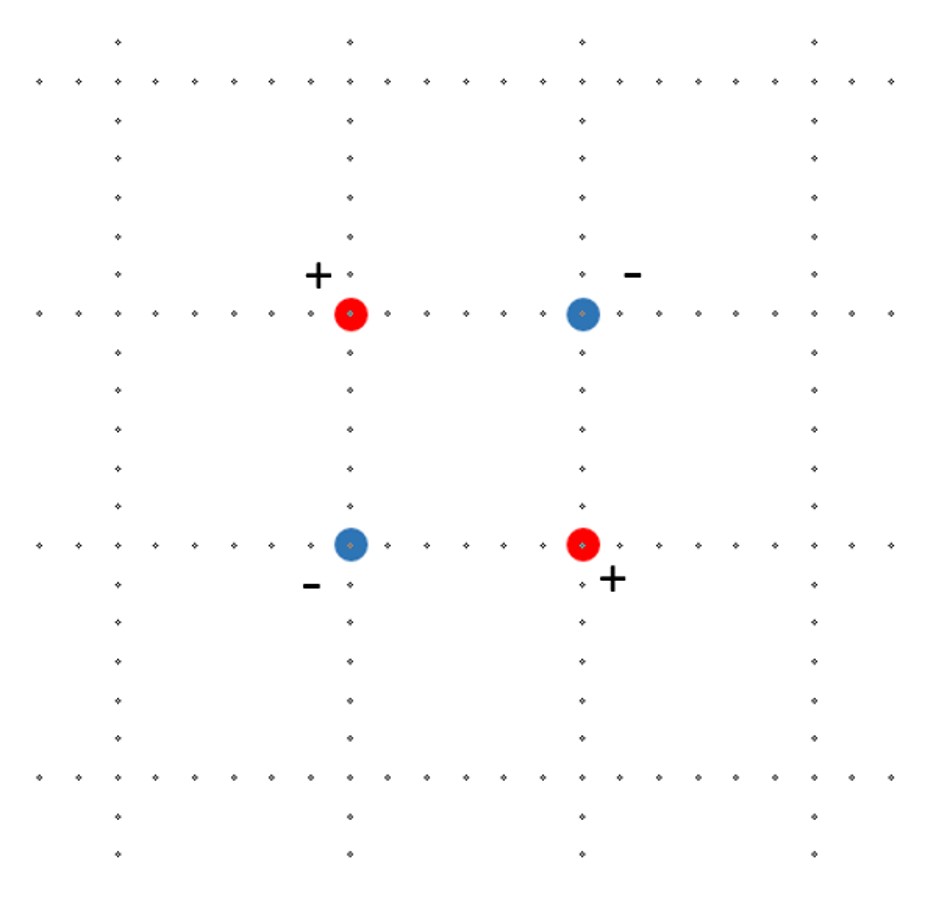}
 \caption{Increasing an off-diagonal component of $E_{ij}$ by 1 creates excitations at the four corners of a plaquette, in a quadrupolar configuration.}
 \label{fig:quadrupole1}
 \end{figure}

\begin{figure}[t!]
 \includegraphics[scale=0.4]{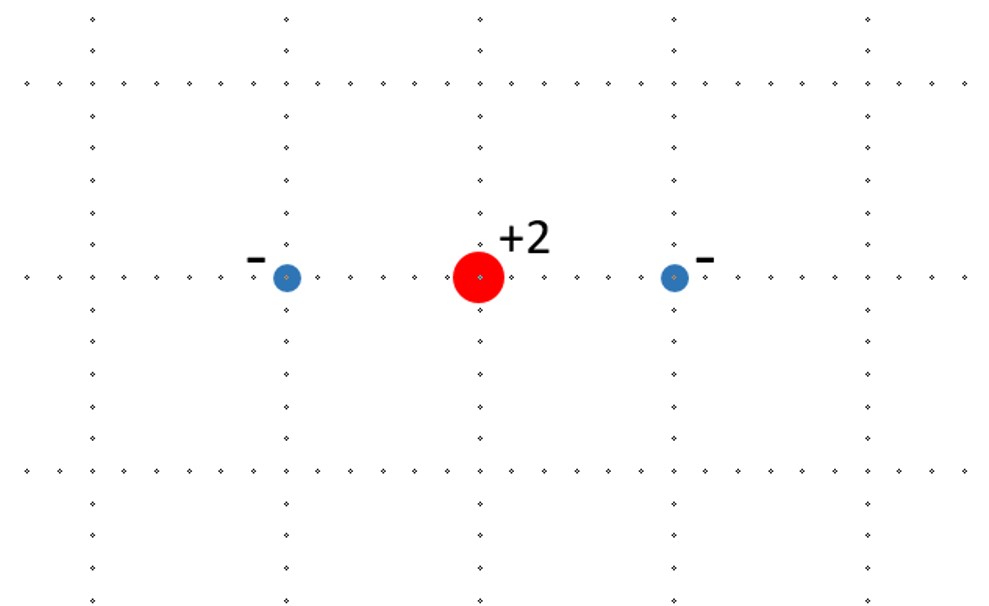}
 \caption{Increasing a diagonal component of $E_{ij}$ by 1 creates three excitations in a line, again with vanishing charge and dipole moment.}
 \label{fig:quadrupole2}
 \end{figure}

However, interestingly, this rank 2 theory has an additional dipolar conservation law:
\begin{equation}
\int \vec{x} \rho = \int x^k \partial_i\partial_j E^{ij} = -\int \partial_j E^{kj} = 0
\end{equation}
where we have integrated by parts in the middle step.\cite{foot5}  (The choice of origin for $\vec{x}$ is arbitrary, since the system is neutral.)  In this theory, therefore, any creation/annihilation operation must not only respect the neutrality of the system, but also the vanishing of its dipole moment.  As a concrete example, take the lattice model discussed in Reference \onlinecite{cenke2}, where the diagonal components ($E_{xx}$, $E_{yy}$, and $E_{zz}$) live on the vertices of a cubic lattice, and the off-diagonal components ($E_{xy}$, $E_{xz}$, and $E_{yz}$) live on the faces, with all components taking integer values.  The basic creation and annihilation operators can be found by examining the effect of changing one component of $E$ at a single location by 1 unit.  Doing so leads to two distinct types of creation and annihilation operators, as shown in Figures \ref{fig:quadrupole1} and \ref{fig:quadrupole2}.

The uniting feature of all such operators is that they correspond to quadrupolar configurations of charge, obeying both charge neutrality and vanishing dipole moment.  In fact, it would seem that this quadrupolar principle is the fundamental feature of this model which would allow it to be generalized to other types of lattices besides cubic.  Putting rotors on the vertices and faces of a cubic lattice allowed for the simplest lattice regularization, since there were effectively six degrees of freedom at each location, corresponding to the six degrees of freedom of a 3$\times$3 symmetric tensor.  Similarly, the simplest lattice regularization of a rank 1 $U(1)$ gauge theory would be on the links of a cubic lattice, giving us three degrees of freedom per site.  Nevertheless, the rank 1 theory can be defined on any lattice, with the fundamental feature being that the electric strings live on links, and these electric strings can essentially be thought of as tiny dipoles.  From this perspective, the rank 2 generalization is quite natural.  Whereas in the rank 1 case microscopic bosonic degrees of freedom were mapped onto dipole creation operators, in a rank 2 theory they map onto quadrupolar creation operators.  We will see that a similar story holds for other higher rank gauge theories.
 
The consequences of this extra conservation law for the motion of individual charges are quite dramatic.  First of all, let us examine an isolated point charge, well-separated from any other charges in the system.  (This situation can indeed be created, for example by a ``membrane" operator, formed by applying the operator of Figure \ref{fig:quadrupole1} on every plaquette of a two-dimensional surface.  Isolated charges will naturally occur at the corners of this membrane.)  We then wish to move such an isolated charge to an adjacent site by means of a local operator.  Upon examining our toolbox, it is obvious that there is no local operator that can hop the particle without simultaneously creating other excitations.  This is a simple consequence of the conservation of dipole moment.  Moving the particle by itself changes the global dipole moment of the system and must be compensated by the creation/destruction/motion of other particles.  Such a particle excitation, which cannot move without the creation of extra excitations, was dubbed a ``fracton" in Reference \onlinecite{fracton1}, and we will stick with that convention.  The difference between the present case and the previously studied fracton models is that the models here also possess gapless gauge modes (which propagate normally).

It should be noted that not all excitations in the system have this fracton property.  Whereas isolated point charges are fractons, a dipolar bound state of two opposite charges will be freely hopping.  The operators seen in Figures \ref{fig:quadrupole1} and \ref{fig:quadrupole2} represent the transverse and longitudinal hops of these dipoles, so dipolar excitations can freely propagate in any direction in space.

\subsection{Vector Charge Theory}

There is another important class of rank 2 symmetric $U(1)$ spin liquids, which has related properties, but with important differences.  In this theory, we take the Gauss's law constraint on the ground state to be $\partial_i E^{ij} = 0$, corresponding to the gauge transformation $A_{ij}\rightarrow A_{ij} + \partial_i \lambda_j + \partial_j \lambda_i$ for arbitrary vector $\lambda_i$.  As in the previous case, we regard the violations of the ground state Gauss's law as the particle states of the theory, defining a vector-valued charge as:
\begin{equation}
\rho^j = \partial_i E^{ij}
\end{equation}
Once again, we first identify the various conservation laws of the system.  We obviously have the natural analogue of charge neutrality in the system:
\begin{equation}
\int \vec{\rho} = \int \partial_i E^{ij} = 0
\end{equation}
We can also consider the moment of this vector charge around an arbitrary origin:
\begin{equation}
\int \vec{x}\times \vec{\rho} = \int \epsilon_{ijk} x^j \partial_n E^{nk} = -\int \epsilon_{ijk}E^{jk} = 0
\end{equation}
where we have once again integrated by parts, and the last step follows since $E^{ij}$ is a symmetric tensor.  If we heuristically regard $\vec{\rho}$ as a ``momentum density", then the two conservation laws here represent the conservation of linear and angular momentum.  And just as any particle creation/annihilation operator in the scalar case needed to respect the neutrality and dipolar conservation laws, any creation/annihilation operator in this theory must respect both linear and angular momentum conservation.

\begin{figure}[b!]
 \includegraphics[scale=0.45]{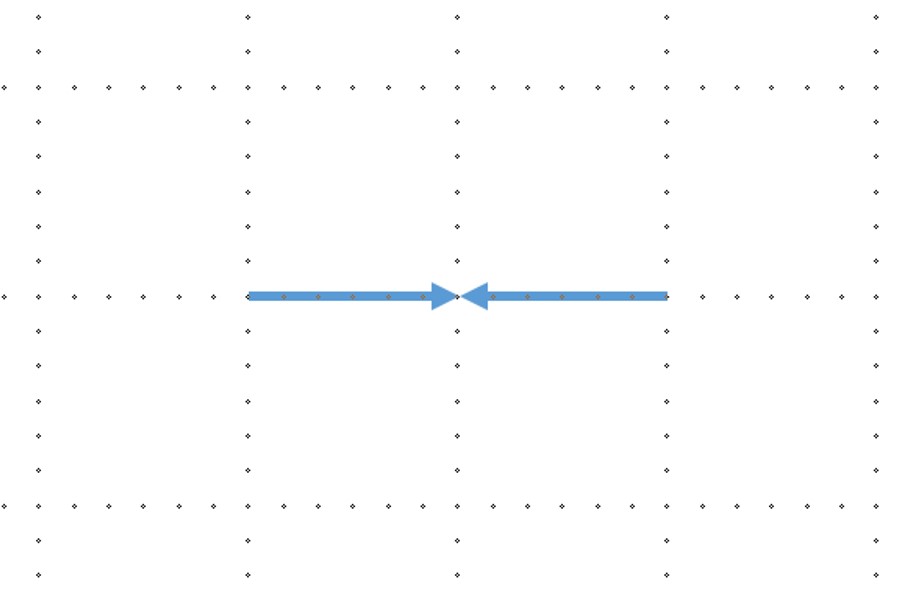}
 \caption{Increasing a diagonal component of $E_{ij}$ by 1 creates two oppositely directed vector charges on adjacent collinear links.}
 \label{fig:vector1}
 \end{figure}

 \begin{figure}[b!]
 \includegraphics[scale=0.35]{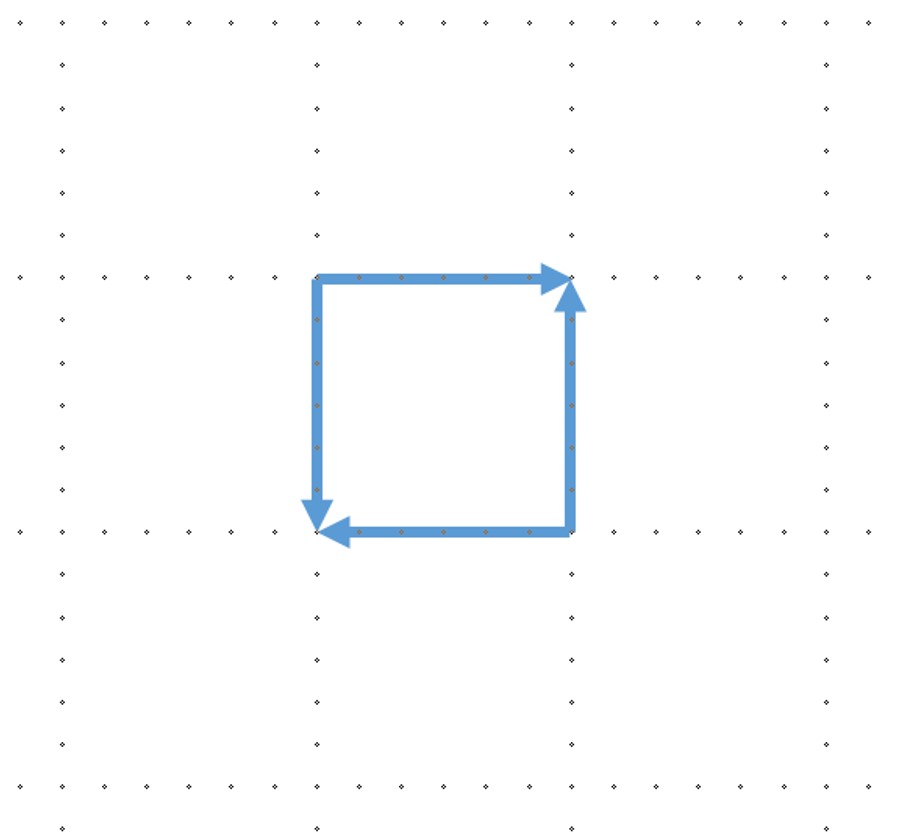}
 \caption{Increasing an off-diagonal component of $E_{ij}$ by 1 creates a ``loop" of vector charge around a plaquette, in such a way that both linear and angular momentum are conserved.}
 \label{fig:vector2}
 \end{figure}

As a concrete example, let us consider the operators present in the rank 2 vector charge model considered in Reference \onlinecite{alex}.  Once again, diagonal components live on vertices of a cubic lattice and off-diagonal components live on the faces.  The vector charges live naturally on the links of the lattice.  There are two sorts of local creation/annihilation operators, as depicted in Figures \ref{fig:vector1} and \ref{fig:vector2}.  One creates opposite vector charges on adjacent collinear links.  The other creates vector charges in a sort of loop around a plaquette.  However, it is important to note that the ``loop" changes orientation at every vertex, in order to respect the conservation laws.  (A unidirectional loop would violate angular momentum conservation.)  The first of these operators, acting on collinear links, can be thought of as a longitudinal hopping operator for the vector charge, allowing each particle to hop freely along the direction of its charge.  However, it is easy to see that an analogous transverse hopping operator is disallowed, since it would violate conservation of angular momentum.\cite{foot6}  We can regard the plaquette operator as generating a transverse hop, but only at the cost of introducing two new excitations on perpendicular links to compensate for the change in angular momentum.  Therefore, the vector charges are ``one-dimensional particles."  They are restricted to motion only along their charge vector, being effectively ``confined" in the transverse direction due to the need to create extra particles.  Note that one-dimensional particles are not unique to the $x$ direction or $y$ direction in Figures \ref{fig:vector1} and \ref{fig:vector2}.  A bound state of an $x$ charge and a $y$ charge will be freely hopping in the $(1,1,0)$ direction, with operators like Figure \ref{fig:vector2} generating the hops of bound states.
 
Once again, the principles identified here provide for a natural way to generalize the model to any lattice, beyond the cubic models considered in previous work \cite{alex,cenke2}.  The principle is to engineer a Gauss's law which maps the local bosonic degrees of freedom onto creation/annihilation operators on links which are consistent with conservation of linear and angular momentum.  Since the vector charges naturally live on links, the gauge variable will naturally live either on faces, sites, or both.  The vector charges will be able to freely hop only along one-dimensional subspaces.

\subsection{Traceless Rank 2 Theories}

We have so far discussed two different forms of gauge constraints, $\partial_i\partial_j E^{ij} = 0$ and $\partial_i E^{ij} = 0$.  If we had the second constraint, then the first is automatic, so there is no sense in combining these two gauge constraints.  The only additional thing we can add on top is a tracelessness constraint, $E^i_{\,\,i} = 0$.  There is actually a subtle distinction regarding whether we treat this constraint as being breakable or not.  In all previous Gauss's law constraints, derivatives were present, and allowing for gauge charges was necessary to preserve the tensor product structure of the Hilbert space (see Reference \onlinecite{me} for further discussion).  However, for a local constraint such as the trace condition, there are two options.  We can allow the constraint to be broken, in which case there are trace charges in the theory, $E^i_{\,\,i} = n$, for integer $n$.  Alternatively, we are free to stipulate that our microscopic variables were traceless and say that $E^i_{\,\,i} = 0$ identically in the Hilbert space.  We shall focus on this latter case, since it leads to more novel physics.  Trace charges would only add in extra local scalar degrees of freedom which could modify the energetics of our system, but otherwise not do much harm.

The tracelessness condition looks fairly benign at first glance.  Its only effect is to even further constrain the particles with an additional conservation law.  For example, consider the scalar charge theory, in which case the extra conservation law is:
\begin{equation}
\int x^2\rho = \int x^kx_k\partial_i\partial_j E^{ij} = 2\int E^i_{\,\,i} = 0
\end{equation}
It can readily be checked that this conservation law rules out linear quadrupole operators such as that in Figure \ref{fig:quadrupole2}, only allowing square quadrupoles such as that in Figure \ref{fig:quadrupole1} and various bound states of the linear quadrupoles.  The fractons remain fractonic, so not much changes in the charge sector.  However, the neutral dipolar bound state, which was formerly freely hopping, now becomes a two-dimensional particle, as its longitudinal hopping operator has been removed from the theory.

However, the tracelessness condition has a much more dramatic effect on the vector charge theory.  Without tracelessness, the two conservation laws of this theory were conservation of ``linear momentum" and ``angular momentum."  In the presence of the tracelessness constraint, there is yet another conservation law:
\begin{equation}
\int \vec{x}\cdot \vec{\rho} = \int x^j\partial_i E^{ij} = -\int E^i_{\,\,i} = 0
\end{equation}
In the presence of this extra conservation law, the vector charges become fully fractonic in nature.  Longitudinal hops, formerly allowed in the traceful theory, are now in violation of the new conservation law.  The formerly one-dimensional particles have been turned into fractons via this extra gauge condition.  As a general principle, adding in additional gauge constraints can only lead to an increase in the number of conservation laws of the system, thereby putting extra constraints on particle motion.

\subsection{``Mixed" Gauge Constraints}

Strictly speaking, there is still one more layer of complication which can be thrown into these theories, which we shall mention only in passing.  Until now, we have always taken the gauge transformation to be on $A$, while the constraint is on $E$, as seems usual.  One could also think up theories in which $A$ and $E$ switch roles, which is probably not dramatically different, resulting in a dual description of the same phases.  However, we could also consider a ``mixed" gauge theory, where constraints are simultaneously imposed on both $A$ and $E$, so long as care is taken to ensure that these constraints commute.  For example, the gauge transformations and constraints of the emergent gravity model in Reference \onlinecite{gu} are effectively:
\begin{equation}
A_{ij}\rightarrow A_{ij} + \partial_i\lambda_j + \partial_j \lambda_i\,\,\,\,\,\,\,\,\,\,\,\,\,\,\,\,\,\,\,\,\,\,\,\,\,\,\,\,\,\,\,\,\,\partial_i E^{ij} = 0
\label{gauge1}
\end{equation}
\begin{equation}
E_{ij} \rightarrow E_{ij} + (\delta_{ij}\partial^2 -\partial_i\partial_j)\phi \,\,\,\,\,\,\,\,\,\,\,\,\,\,\,\,\,\, (\delta_{ij}\partial^2 - \partial_i\partial_j)A^{ij} = 0
\label{gauge2}
\end{equation}
for arbitrary gauge parameters $\lambda_i$ and $\phi$.  It is easy to verify that the two constraints commute, simply by examining the effect of the opposite gauge transformation on each constraint.  We therefore see that mixed gauge constraints are in principle possible, though it is unclear when such theories are physically reasonable.  For example, in the rank 1 case, we could have tried to simultaneously impose both the normal Gauss's law, $\partial_iE^i = 0$, and also a curl constraint on $A$ as $\nabla \times\vec{A} = \vec{B} = 0$.  However, imposing this extra magnetic flux constraint takes us to a ``sick" limit of the $U(1)$ gauge theory, where the speed of light has gone to infinity, so there is no longer a sensible gapless gauge mode.  As shown in Reference \onlinecite{gu}, it is possible in certain circumstances for such mixed gauge constraints to lead to reasonable theories.  It is unclear under what conditions these mixed gauge constraints will leave us with a reasonable theory, as opposed to some ``sick" theory with infinite velocities.

For these theories, we have both an ``electric" gauge constraint and a ``magnetic" gauge constraint, both of which have corresponding particles.  It is important to note that these particles are not the magnetic defects of the theory.  On top of these magnetic gauge constraint particles, the theory can still have magnetic defects arising from the compact nature of the gauge variable.  These mixed gauge theories therefore have a particularly complicated particle spectrum.

\subsection{Magnetic Defects and Stability}

Up to this point, we have technically only been speaking in terms of the ``electric" particles of the theory, $i.e.$ those coming from violations of the Gauss's law constraint.  However, we must also consider the magnetic defects of the theory.  We argued earlier that, in the rank 1 $U(1)$ gauge theory, the magnetic defects have basically the same structure as the electric charges, as determined by an electric-magnetic duality.  Similar stories hold in many of the higher rank theories.  For example, take the rank 2 vector charge theory, with low-energy gauge constraint $\partial_i E^{ij} = 0$.  Writing down the gauge invariant quantity with the fewest number of derivatives, the magnetic field in this case was shown\cite{alex} to also be given by a rank 2 symmetric tensor:
\begin{equation}
B_{ij} = \epsilon_{iab}\epsilon_{jcd}\partial_a\partial_c A_{bd}
\end{equation}
which appears to obey $\partial_i B^{ij} = 0$.  However, once again, due to the compactness of $A$, we can have $\partial_i B^{ij} = 2\pi n^j$, for integer valued vector $n^j$.  By the definition of $n^j$, we can apply the same arguments used on the vector electric charge to show that these vector defects obey conservation of both ``linear momentum" and ``angular momentum," so they are also one-dimensional particles.  As in the conventional $U(1)$ spin liquid, this could have been determined by electric-magnetic duality.  As shown in Reference \onlinecite{alex}, one can use the low-energy gauge constraint to write $E^{ij} = \epsilon^{iab}\epsilon^{jcd}\partial_a\partial_c \tilde{A}_{bd}$ for a dual gauge field $\tilde{A}_{ij}$, which effectively swaps the role of $E$ and $B$ in the theory, so the magnetic charges must have the same structure as the electric charges.  For any theory with such duality, it is automatically the case that the magnetic charges have the same subdimensional behavior as the electric charges, which is a nice result.  In particular, this immediately tells us that the magnetic defects are excitations, not instantons.  This allows us to consider a regime in which the magnetic excitations are gapped and are irrelevant to the low-energy physics, providing us with a stable phase of matter.

There are also some phases which are not self-dual.  For example, consider the scalar charge theory, without any trace conditions.  One can show that the lowest order magnetic field tensor which can be written down has the non-symmetric form\cite{foot9} $B_{ij} = \epsilon_{iab}\partial^a A^b_{\,\,j}$.  Despite the lack of self-duality, the theory still has magnetic monopole excitations of the form $\partial_iB^{ij} = 2\pi n^j$, which one can show are 2-dimensional particles, moving transversely to their charge vector.  The presence of pointlike magnetic monopole excitations seems to be quite universal to these symmetric tensor phases, which indicates that they are all stable phases of matter.

We can now say that these phases are stable against confinement by instanton proliferation, but one might worry that one deconfined phase could be unstable to a different deconfined phase.  For example, suppose we take the gauge constraint to be $\partial_i\partial_j E^{ij} = 0$.  Our ideal Hamiltonian will have the schematic form:
\begin{equation}
H = E^2 + B^2 + U(\partial_i\partial_j E^{ij})^2
\end{equation}
where $B$ is the appropriate magnetic tensor.  However, one might complain that we could throw in a term corresponding to the lower order gauge constraint of the vector charge theory:
\begin{equation}
H = E^2 + B^2 + U(\partial_i\partial_j E^{ij})^2 + U'(\partial_i E^{ij})^2
\label{thisone}
\end{equation}
The $U'$ term naively looks more relevant, since it has fewer derivatives.  If this were indeed true, then the scalar charge rank 2 theory would flow towards the vector charge rank 2 theory.  However, this is not the case.  We cannot view the last two terms of Equation \ref{thisone} on equal footing, since the particular $B$ here commutes only with the first gauge constraint, not the second.  The $B$ term and the $U'$ term are in competition.  At small $U'$, the magnetic term will win out and the $U'$ term is irrelevant.  Another way to see this is to note that the addition of the $U'$ term only modifies the dispersion relation from $\omega^2-k^4 = 0$ to $\omega^2 + U'\omega^2k^2 - k^4 = 0$, which is an irrelevant change at low energies.  Of course, at large enough $U'$ (far away from the fixed point), the vector gauge constraint will take over.  But for small perturbations, the scalar charge fixed point is a locally stable one against the vector perturbation.  Similar arguments hold for the stability of any of the gauge constraint combinations against any other.  The $B$ term for each theory is constructed specifically for that gauge constraint and will destroy any other gauge constraint added in perturbatively.

\section{Electrostatic Confinement}

(Special thanks are due in this section to Senthil Todadri, who first pointed out that electrostatic energy may be an issue in these models.)

Before moving ahead with these models, we need to take a close look at the energy needed to create well-separated fractons.  Besides the energy $U$ representing the mass of the particles, we must also consider the energy stored in the ``electric field" $E_{ij}$ of a static point charge.  (More correctly, we are looking at the expectation value $\langle E_{ij}\rangle$, but we shall simply write $E_{ij}$ to avoid clutter.)  Suppose we have a static delta function source for our Gauss's law, representing an isolated point charge.  In the vector charge theory, which has one derivative in Gauss's law, we have:
\begin{equation}
\partial_i E^{ij} = v^j \delta(r)
\end{equation}
for some constant vector $v^j$.  Fourier transforming gives us:
\begin{equation}
q_i \tilde{E}^{ij} = v^j
\end{equation}
The scaling of our electric field then behaves as:
\begin{equation}
\tilde{E}^{ij}(q) \sim \frac{1}{q}
\end{equation}
\begin{equation}
E^{ij}(r) \sim \int d^3q \,\tilde{E}^{ij}(q) e^{i\vec{q}\cdot\vec{r}} \sim \frac{1}{r^2}
\end{equation}
like a normal Coulomb field.  The electrostatic energy associated with these single-derivative theories is much the same as in a rank 1 $U(1)$ spin liquid, and no further comments are necessary.

On the other hand, consider a delta function source for the scalar charge theory:
\begin{equation}
\partial_i\partial_j E^{ij} = \delta(r)
\end{equation}
The scaling of the electric field now behaves as:
\begin{equation}
\tilde{E}^{ij}(q) \sim \frac{1}{q^2}
\end{equation}
\begin{equation}
E^{ij}(r) = \int d^3q\, \tilde{E}^{ij}(q) e^{i\vec{q}\cdot\vec{r}} \sim \frac{1}{r}
\end{equation}
so the electric field in this case only falls off as $1/r$, as opposed to the $1/r^2$ behavior of the usual three-dimensional Coulomb field.  This is quite a significant difference, since the energy stored in the field is given by:
\begin{equation}
\int d^3r E^2 \sim \int dr\, r^2 \frac{1}{r^2} \sim R
\end{equation}
where $R$ is a large-distance cut-off.  In the usual Coulomb field, $1/r^2$, the integral converges at large distances, indicating a finite amount of energy stored in the electric field of a point particle.  (Short-distance divergences are of course cut off at a finite value in any condensed matter context.)  For the rank 2 scalar charge theory, however, the electrostatic energy of a truly isolated point charge is actually infinite (cut off by the system size).  And if a group of particles with zero net charge and dipole moment is separated from each other by a distance of order $R$, then the natural energy scale of this system is of order $R$.  We have therefore found that a configuration with particles separated by distance $R$ requires an energy linear in the separation.  This is the standard metric by which one normally measures confinement.

However, it is important to note that this is a different sort of ``confinement" which is much milder than the usual notion.  Usually, one speaks in terms of the interaction between particles.  If a linear potential, $V(R)\sim R$, exists between two particles, then there is an attractive force which is independent of distance, always seeking to recombine the particles into the vacuum, indicating that the particles never truly become independent and are thus confined.  However, in the fracton case, the electric field of a point particle dies off as $1/r$, so at large separation, the particle's do not feel each other's fields.  The linear energy associated with separating particles does not arise from one particle working against the field of the other, but rather from the large energy associated with building up each individual charge's electrostatic field.

Normally, these two notions are equivalent to each other.  For example, a standard undergraduate electromagnetism problem\cite{griffiths} is to calculate the electrostatic energy of a uniformly charged ball.  One way to do the problem is to examine the energy stored in the final electric field configuration.  Another way to do it is to imagine building up the ball incrementally and calculate the work done in bringing each charge into the ball against the field of the charges already present.  In the present problem, however, the two descriptions are inequivalent, as we just found.  The resolution to this seeming paradox is that there simply is no notion of ``force" (or even equations of motion) when dealing with fractons.  Since fractons cannot hop, there is no sense in which one can do work to move a particle against the field.  Any treatment involving incremental energy changes would need to account for the creation of extra particles along the way, which seems like a complicated task.  It is much simpler to just look at the final electrostatic energy, which unambiguously yields the answer.

The fractons in this theory are somewhat ``confined," in the sense that it requires a macroscopically large energy to acquire well-separated fractons, but many aspects of traditional Polyakov confinement do not carry over.  First of all, the gauge field is not gapped out (none of the present issues affect the stability arguments of Reference \onlinecite{alex}), so this ``confinement" does not destroy the Coulomb phase.  Furthermore, once the large energy cost has been paid to create well-separated fractons, there will be no ``restoring force" on an isolated fracton which seeks to recombine it with its neutralizing partners into the vacuum.  A fracton cannot move towards its neutralizing partners without creating extra particles, the cost of which will energetically outweigh any marginal decrease in the electrostatic energy.

In this sense, it is still perfectly reasonable to speak of these fractons as well-defined excitations of the system.  The isolated fracton state is stable, just energetically costly, much like vortices in a two-dimensional superfluid.  A configuration with well-separated fractons is too energetically costly to be created via thermal fluctuations, so they will have little effect on low-energy thermal properties of the system.  But if one prepared a system with well-separated fractons, then these fractons would continue to exist as stable excitations of the system.  We will give this new milder version of confinement the name ``electrostatic confinement."  For this rank 2 theory, electrostatic confinement actually has more in common with a conventional deconfined phase (well-defined particle excitations and a gapless gauge field) and the main effect of the confinement is just to inhibit the fractons from being generated by thermal fluctuations.

Despite its benign behavior in rank 2 theories, we will later face examples at rank 3 and higher where electrostatic confinement is more severe.  In the traditional Gauss's law with one derivative, we had $E\sim 1/r^2$.  In the case with two derivatives, we have $E\sim 1/r$.  In a theory with three derivatives in Gauss's law, such as $\partial_i\partial_j\partial_k E^{ijk} = 0$, we would have $E\sim \log(r)$, which actually increases with distance.  For higher derivatives, the growth is even faster.  In such a theory, we face the additional problem that, not only is the total electrostatic energy divergent, but the energy \emph{density} at large distances diverges as well.  At some critical distance, the energy density would be large enough to start nucleating new particles out of the vacuum.  This would then place a fundamental limitation on the isolatability of an individual fracton.  Thus, in theories with three derivatives or more in Gauss's law, isolated fractons do not appear as stable excitations at long-distances.  Such theories will still have protected gaplessness, but all particles will only occur in neutral bound states.  The neutral bound states can actually still be subdimensional in certain cases, so these theories are far from trivial, but the charge sector will be totally destroyed.

\section{Generalization to Higher Rank}

We can very straightforwardly generalize the principles of the rank 2 theories to higher rank cases.  At higher ranks, there are a wider variety of available Gauss's laws, as discussed in Reference \onlinecite{alex}.  All of these theories can be analyzed from the same perspective as the rank 2 case: identify all charge conservation laws, then consider all local creation/annihilation operators respecting those conservation laws.  For one- or two-derivative Gauss's laws, we can then immediately identify all subdimensional behavior in the theory.  As noted earlier, if there are three or more derivatives present in the Gauss's law, then the fundamental charge sector will be destroyed by electrostatic confinement, even though the gapless gauge mode is still protected.  Neutral bound states will still exist as well-defined excitations, and may still be subdimensional.  Such theories are therefore not uninteresting, but are slightly less interesting than the one- and two-derivative cases, which have well-defined charge sectors.  We will focus on these two cases for the sake of brevity.  Any other specific case can be studied via the same principles.

In the first case, we take our Gauss's law to have two derivatives, $\partial_{i_1}\partial_{i_2} E^{i_1...i_n} = \rho^{i_3...i_n}$.  (We can also allow the other indices to be contracted with each other in various ways.  Contracting internal indices of $E$ just lowers its effective rank, which would allow us to study it by previous methods.)  Since there are two derivatives, it is easy to see that the first two charge moments vanish:
\begin{equation}
\int \rho^{i_3...i_n} = 0
\end{equation}
\begin{equation}
\int x^j \rho^{i_3...i_n} = 0
\end{equation}
after performing an integration by parts in the second equation.  When coupled together, these equations imply that isolated charges are fully immobile.  Such a tensor charge cannot hop to an adjacent site while preserving both charge and first moment.  Thus, two-derivative models are all fracton theories.  Just as in the rank 2 case, two-derivative theories at any rank will have a $1/r$ electric field for isolated charges.  Electrostatic confinement will then prevent the particles from being generated via thermal fluctuations, but they will still exist as stable states.  (Higher derivative theories would also have all charged particles be fractonic, but these charges would be destabilized by electrostatic confinement.)

The other case occurs when there is one derivative in Gauss's law, $\partial_{i_1} E^{i_1...i_n} = \rho^{i_2...i_n}$.  We obviously have charge conservation:
\begin{equation}
\int \rho^{i_2...i_n} = 0
\end{equation}
But we no longer have the vanishing of a generic first moment:
\begin{equation}
\int x^j \rho^{i_2...i_n} = \int x^j \partial_{i_1} E^{i_1...i_n} = -\int E^{ji_2...i_n} \neq 0
\end{equation}
However, by taking advantage of the symmetry of $E$, a fully symmetric tensor, we can easily construct a contracted first moment which does obey a conservation law:
\begin{equation}
\int \epsilon^{nji_2} x_j \rho_{i_2...i_n} = -\int \epsilon^{nji_2} E_{ji_2...i_n} = 0
\end{equation}
by the symmetry of $E$.  In the rank 2 case, the analogous conservation law was conservation of angular moment, which still allowed the charges to freely hop along the direction of their charge vector.  However, for rank higher than 2, the charges have at least two indices.  In this case, the conservation law is more constraining.  For a charge $\rho_{i_2...i_n}$ to be freely hopping in direction $v_j$, we require $\epsilon^{nji_2}v_j\rho_{i_2...i_n} = 0$ for all $n$, so that the moment does not change with hops in the $v_j$ direction.  For this to be true for arbitrary $n$ (and considering the index symmetry of $\rho$), $\rho$ cannot have components in any index direction except the $v_j$ direction.  In other words, a charge can only propagate in direction $v_i$ if the charge tensor has the specific form $\rho_{i_2...i_n} = v_{i_2}v_{i_3}...v_{i_n}$.  A generic tensor will have no directions of free propagation and will therefore be a fracton.  This case therefore has a mix of both one-dimensional particles and fractons (with most charges being fractons).

\section{Conclusion}

We have shown that spin liquids described by higher rank $U(1)$ gauge theories \cite{alex} lead naturally to subdimensional particle excitations, as in fracton models \cite{fracton1,fracton2}.  The theories with scalar charge are fractonic, with particles unable to hop in any direction without generating additional excitations.  The vector charge theories can be either fractonic or have one-dimensional excitations.  These models provide a natural gapless generalization of the original discrete fracton models.  Furthermore, it seems like tensor gauge theories may provide a more general framework for understanding and classifying phases with subdimensional particles.  There is much work to be done on this front, and exciting new theoretical questions are waiting to be answered.  Do all of the previously discovered fracton models fit into the tensor gauge theory framework?  How do these formulations relate to previous work relying on algebraic geometry?  How does one obtain the fractal structure of Haah's code?  Also, recent work has indicated that discrete fracton models can be obtained from a layer construction method, where layers of two-dimensional topologically ordered systems are strongly coupled together\cite{sagarlayer,hanlayer}.  It will be interesting to see if the $U(1)$ fracton models can be obtained from similar constructions.

There are plenty of even more exotic (and perhaps less well-defined) questions.  Can we get anything new by considering non-abelian generalizations of these tensor gauge theories?  What is the nature of transitions to the confined phase?  Can subdimensional particles condense?  Can subdimensional particles exist on curved submanifolds?  For example, could we construct a model in which two-dimensional particles are restricted to a sphere instead of a plane?  There are many new avenues to explore.

On a more practical level, how and where can we observe such subdimensional particles in the lab?  As to how, we note that the conservation laws of these theories will have dramatic consequences for thermalization properties of these phases\cite{abhinav}.  Any energy injected into the particle modes of these systems tends to be localized to a particular subspace (or to one particular point, in the case of fractons).  Thus, these systems have severe obstructions to thermalization and will have unusual thermal transport properties.  Furthermore, this localization exists in the clean limit, without the disorder that is often needed in the context of many-body localization, which may serve as a way to distinguish subdimensional particles from more conventional localization effects.

The specific lattice models of Reference \onlinecite{alex} are theoretically well-established to be in the higher rank $U(1)$ spin liquid phase.  These models apply most directly to a system of quantum rotors, which is essentially equivalent to a large $S$ limit of spins.  Such a quantum rotor system could possibly be engineered directly in ultra-cold atom systems, which philosophically is enough to establish the reality of the phases described here.  But there is every reason to believe that these phases are realizable in Mott-insulating solid-state systems as well.  A finite $S$ system, even a simple spin-$1/2$ system, can flow towards a rotor description under the renormalization group\cite{balents}.  The precise choice of microscopics is unimportant.  The key ingredient for these models is the Gauss's law, which corresponds to some strong frustrating interaction of the underlying spins.  The precise form of the geometric frustration will necessarily be more complicated in these models than in the rank 1 case, since spin flips should create more than just two particles.  But such geometries are definitely accessible, such as the face-centered cubic geometry of the models in Reference \onlinecite{alex}.  It remains to be seen if any fcc material will have the precise interaction structure necessary to produce higher rank $U(1)$ spin liquids, or if a more complicated geometry is needed.  Either way, there is every reason to be optimistic that higher rank $U(1)$ spin liquids are experimentally accessible.

\section*{Acknowledgements}

Numerous thanks are due to my advisor, Senthil Todadri, first and foremost for calling my attention to the large electrostatic energy of these models.  He is also due thanks for many insightful discussions, for financial support throughout, and for allowing me sufficient freedom to follow my own interests.  I would also particularly like to thank Liujun Zou for numerous helpful discussions during the early stages of this project.  Thanks are due to Sagar Vijay, Jeongwan Haah, and Liang Fu for sharing their expertise on fracton models.  I would also like to thank Yahui Zhang, Lucile Savary, and Samuel Lederer for useful discussions.  This work was supported by NSF DMR-1305741.

\section*{Appendix:  What is a Tensor?}

\begin{figure}[t!]
 \includegraphics[scale=0.35]{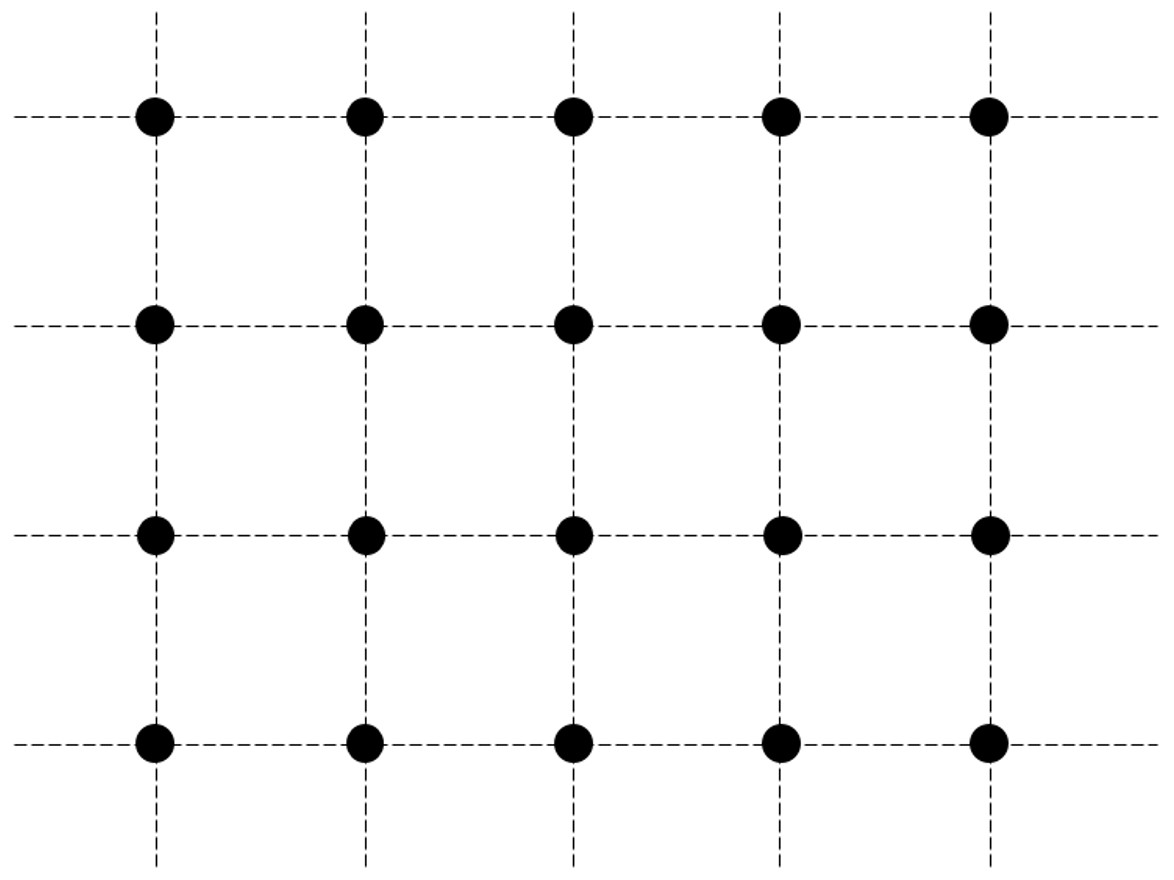}
 \caption{We can regard the spins of this system as living on the sites of a square lattice, in a scalar representation.}
 \label{fig:square}
 \end{figure}

 \begin{figure}[t!]
 \includegraphics[scale=0.35]{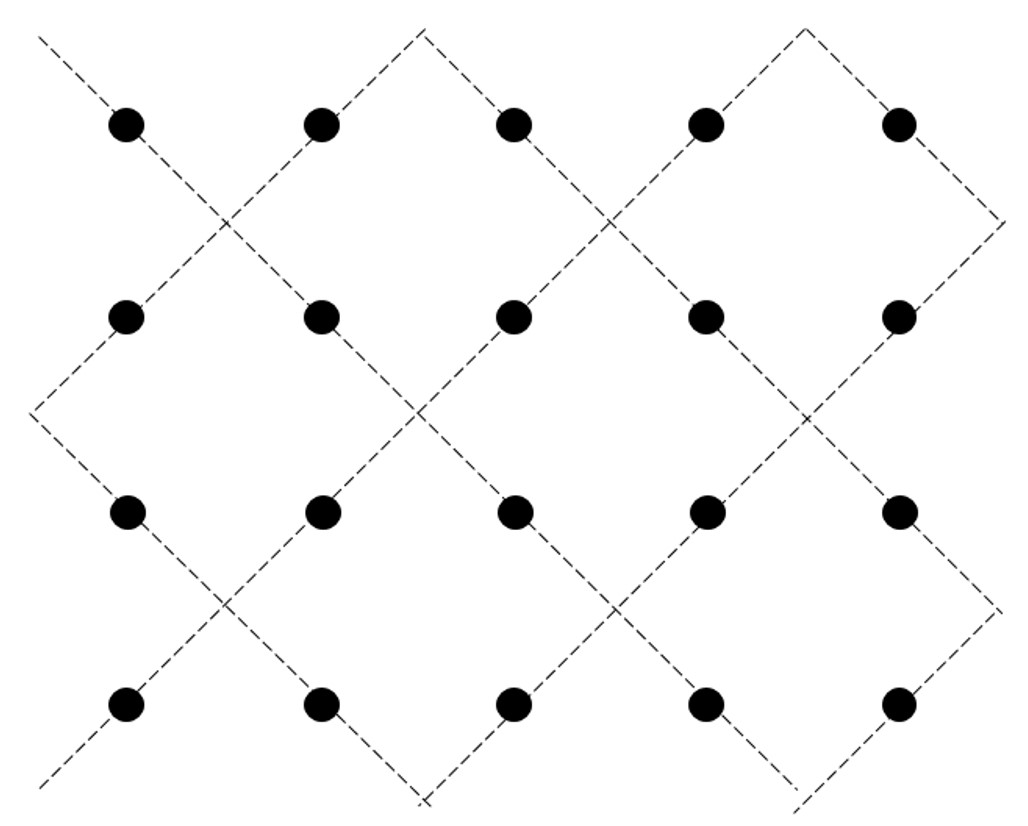}
 \caption{We could regard the same system of spins as living on the links of a tilted square lattice, in a vector representation.}
 \label{fig:square2}
 \end{figure}

In the main text, we have considered models in which the fundamental bosonic degrees of freedom map onto a ``tensor" structure.  This is essentially just a rewriting of our Hilbert space.  For example, consider spins on the square lattice in Figure \ref{fig:square}.  Under a $\pi/2$ rotation, the lattice returns to itself, and every spin ends up at an equivalent location.  The fundamental variables are not endowed with any extra sense of directionality (not counting any internal spin direction), so this is effectively a ``scalar" representation of the spin model, as would be appropriate to describe a ferromagnet, for example.  On the other hand, let us now take the exact same spin system, but represent the spins as living on the links of a tilted square lattice, as seen in Figure \ref{fig:square2}.  An excitation of the spin can then be represented by a string on this link, which can be directed or undirected, depending on the microscopic details.  When we perform a $\pi/2$ rotation, the system still returns to itself.  However, spins which formerly lived on links in the $(1,1)$ direction now live on links in the $(-1,1)$ direction, so the spin variables transform as a vector in this representation, which is the correct representation for conventional vector gauge theories.  Similarly, in a three-dimensional system, one could view the spins as living on plaquettes, leading naturally to an antisymmetric tensor representation.  One can also get a symmetric tensor representation by allowing some spins to live on vertices and others to live on plaquettes, as discussed in the main text.  We therefore see that we can always map any given spin system back and forth between various representations of the lattice symmetry group.  However, obviously the system is not well-described by all of these tensors at the same time.  The question is which representation is convenient for describing the spectrum of the theory.  Depending on the Hamiltonian of the system in question, one representation may be more useful than any other.  Or sometimes there are multiple useful tensor representations, such as a three-dimensional $\mathbb{Z}_2$ spin liquid, which can be described either as a vector or antisymmetric tensor theory.  Typically, such a tensor rewriting of the theory is only useful when there is some local energetic constraint which forces loops to be closed, or surfaces to be closed, or some other convenient geometric constraint.  This will give a ``Gauss's law" for the system, which means the system can be described by a gauge theory.

Tensors are usually defined in terms of their behavior under spatial rotations.  A generic tensor can be decomposed in terms of certain simpler tensors.  For example, a rank 2 tensor can be written as a sum of an antisymmetric tensor, a traceless symmetric tensor, and a scalar trace.  Studying these irreducible tensor theories should then allow us to describe arbitrary tensor gauge theories.  But throughout the present work, and in any typical condensed matter system, we are always working on a lattice, so there is no full rotational symmetry.  However, on any crystalline lattice there still exists some subgroup of the full rotation group which is a symmetry of the system.  We should then technically switch to speaking in terms of irreducible representations of the discrete rotational group, instead of the full rotational group, but the situation should not be dramatically different, just a bit more tedious to classify.  Plus, it is often the case that full rotational symmetry is restored in the low-energy theory, since anisotropy terms are often irrelevant under the renormalization group.

On this basis, we have focused on tensor representations of the full rotation group in this paper.  However, we would like to emphasize, as was noted in Reference \onlinecite{alex}, that this rotational symmetry is not needed to protect the gaplessness of the gauge mode.  This gaplessness is enforced by the gauge invariance ($i.e.$ by the source-free Gauss's law of the ground state), which is enforced by the gapped nature of the particles.  Essentially, the particle gap protects the gauge mode gaplessness.  In turn, the gauge constraint requires a certain structure for the particle sector of the theory, which in the models considered here turns out to be subdimensional.  In this sense, these phases are stable against microscopic perturbations, without having to worry about symmetry protection.  We also note that our rewriting of the fundamental bosonic degrees of freedom in tensor representations did not assume rotational symmetry of the phase.  It merely took advantage of the symmetry of the underlying Hilbert space.  These tensor representations could be used just as well to describe a phase with broken rotational symmetry.


\begin{thebibliography}{9}
\bibitem{foot1}
We here use ``spin" in a generalized sense referring to any bosonic degrees of freedom, including conventional spins (with a specific behavior under rotations), rotors, itinerant bosons, or simply a general discrete-leveled object (qudit).  The highly entangled phases of any of these can be described in a unified framework via deconfined gauge theories.  Even the case of fermionic microscopic degrees of freedom is only mildly different.  ``Long-range entangled phase" may be more accurate, but we stick with ``spin liquid" for now, to keep closer contact with existing literature.
\bibitem{foot2}
The lower critical spatial dimension for deconfinement of a rank $n$ compact $U(1)$ antisymmetric tensor gauge theory is $2+n$.  In a higher-dimensional world, we would have lots of string-theoretic spin liquids to study.
\bibitem{foot3}
It should be noted that $U(1)$ does not seem to represent a ``gauge group" of the theory, as these theories are outside the conventional Yang-Mills Lie algebra framework.  As we will see, there are no conventional matter fields transforming under a group representation.  We use the terminology $U(1)$ only to indicate the values of the gauge variable, which may not be ideal.  One would expect that a better naming convention will arise from a more formal mathematical classification of these sorts of gauge theories, analogous to the Lie algebra framework for the rank 1 case.  We hope that the sins of the present naming convention are not overly severe.
\bibitem{foot4}
We are referring to conventional confinement here.  Electrostatic confinement will be discussed separately.
\bibitem{foot5}
For a closed manifold, one must introduce a branch cut for $\vec{x}$.  This branch cut can be chosen away from all the charges and causes no complications.  Alternatively, one can work with an open manifold, on which we reach the similar conclusion that dipole moment is conserved in the bulk.
\bibitem{foot6}
In the specific model of Reference \onlinecite{cenke2}, the longitudinal hopping operator is actually twice the charge strength of the ``loop" operator, but this seems to be a non-universal feature and only really serves to redefine the fundamental unit of mobile charge, since strength 1 charges effectively become fractonic.
\bibitem{foot9}
One could try to work with a symmetrized version of this tensor, which is also gauge-invariant, but this ends up being fine-tuned, as one of the gauge modes is tuned to have exactly flat dispersion.  The non-symmetric tensor describes the point at which all gauge modes have the same speed.
\end{thebibliography}
\end{document}